\documentstyle[epsfig,12pt]{article}
\begin{document}
\def\be {\begin{equation}}
\def\ee {\end{equation}}
\newcommand{\rif}[1] {\ref{#1}}
\begin{flushright}
LNF--97/014(P)
\end{flushright}
\centerline{\bf Top Mesons}
\centerline{N. Fabiano\footnote{fabiano@lnf.infn.it}}
\centerline{Perugia University , Via Elce di sotto, Perugia, Italy and}
\centerline{INFN National Laboratories, P.O.Box 13, I00044 Frascati, Italy}
\par\vskip 1.5 cm
\centerline{\bf Abstract}
\par\vskip 5 mm
The possibility of formation for a bound state of a $t$ quark 
and a lighter one is investigated using potential model predictions
and heavy quark effective theory approach.
Resulting estimates for the 1S--2S splitting of the energy levels
are compared to the total
top decay width $\Gamma_t$ .
As for the case of toponium, our conclusions show that the probability of 
formation for \mbox{T--mesons} is negligibly small due to the high top mass 
value.
\newpage
\section{Introduction}

It is conventional wisdom \cite{FADIN} that the top quark has 
no probability of meson 
formation of any kind because the $t$ quark decays in a way that is too fast 
to allow for a single orbit of the bound state.
First calculations based on a relatively light top quark \cite{PRR} have shown
that the observability of $t \overline{t}$ bound states would be possible 
for a narrow window on $m_t$ values. Subsequent estimates \cite{KUHNTOP,NOI}
have demonstrated that there is only a small probability for creation
of a $t \overline{t}$ bound state for higher top masses.
From these results it could be inferred
that even the formation possibility for a meson made out of a single $t$ 
quark and a lighter one is small. In this note we discuss quantitatively the 
last point in light of the present high values for 
top mass \cite{TOPDISCCDF,TOPDISCD0} .
For this purpose we shall make use of calculations from
potential models in QCD in both nonrelativistic and relativistic theories.
The choice of the potentials is driven by considerations on the proper QCD 
scale for this problem.
We will also give some results taken from heavy quark effective theory, based
on an estimate for the inertia parameter $\overline{\Lambda}$ .
The latter method is particularly suitable for mesons containing a very
light quark.

These results will be compared to the decay probability of the top quark, whose
value is determined by $m_t$ , in order to give quantitative answers on the 
possibility of creation for such ``superheavy'' mesons.

The paper is organized as follows: we review the toponium case 
in Sect. 2. 
The potential model approach for the T--mesons, both nonrelativistic and 
relativistic, is presented in Sect. 3. The heavy quark effective theory 
discussion for these mesons is in Sect. 4. Sect. 5 is devoted to the 
discussion and conclusions.

\section{Toponium bound states revisited}
The recent discovery of the top quark by CDF \cite{TOPDISCCDF} and 
\mbox{D\O}  collaboration \cite{TOPDISCD0}, and subsequent measurements
\cite{TOPMESCDF} and  \cite{TOPMESD0} have given large values for the top mass:

\begin{equation}
m_t=175.6 \pm 9.3 \: \mbox{GeV (CDF)} \: ; \; m_t=169 \pm 11 \: \mbox{GeV (D\O)}
\label{eq:topmass}
\end{equation}
Previous top mass values of around 130 GeV  already excluded the possibility 
of formation of toponium bound states \cite{KUHNTOP,NOI} .

This fact can be understood by comparing the time period of the would--be 
bound state with the decay probability of the top quark. For $ t $ mass 
above the $Wb$ threshold, the top decays into a real $W$ and a $b$ 
quark \cite{BIGI,BARGER}.
The single quark decay width of the top quark, with one--loop QCD correction 
and neglecting terms of order $(m_b/m_t)^2$ , is given by \cite{TOPWIDTH} :

\begin{equation}
 \Gamma_t = \frac{m_t^3}{16 \pi v^2} \left [ 1- \left ( \frac{M_W}{m_t}
\right )^2 \right ]^2  \left [ 1+2 \left ( \frac{M_W}{m_t} \right )^2 \right 
] \left [ 1-\frac{2}{3} \frac{\alpha_s}{\pi} \left ( \frac{2 \pi^2}{3} 
- \frac{5}{2} \right ) \right ] 
\label{eq:width}
\end{equation}
The width values 
increase from $438$ MeV (for $m_t=120$ GeV) to $1.02$ GeV ($m_t=160$ GeV), and
$2.22$ GeV for $m_t=200$ GeV .
With such a large width, the lifetime of T--mesons and toponium are 
dominated by single quark decays.

For a bound state, the typical formation time of a hadron is characterized 
by a revolution time driven by strong interactions. Thus, no bound states 
exist, if the revolution time, $t_R=2\pi r/v$, is larger than the 
lifetime of the rotating quarks, $\tau_{{t \overline{t}}}=
1/\Gamma_{{t \overline{t}}}$ \cite{BIGI} .

The value for the width of the toponium system is two times the width of the 
single top quark, since each one could decay in an independent manner: 
$ \Gamma_{t \overline{t}} =2 \Gamma_t $ .
To discuss the bound state, let us begin with  the toponium case through
 a Coulombic two--body potential:

\begin{equation}
V(r)= -\frac{4}{3} \; \frac{ \alpha_s}{r}
\label{eq:coulomb}
\end{equation}
for which analytic solutions exist. Here should also be considered  corrections
from Higgs boson exchange Yukawa type forces \cite{INAZAWA,STRASSLER}  . 
For a top quark with mass less than $\approx 200$ GeV these corrections 
are small, with an attractive potential for the quark--antiquark singlet 
state, and they amount to no more than $10 \%$ of the Coulomb term.

\begin{figure}[p]
\begin{center}
\leavevmode
\mbox{\epsfig{file=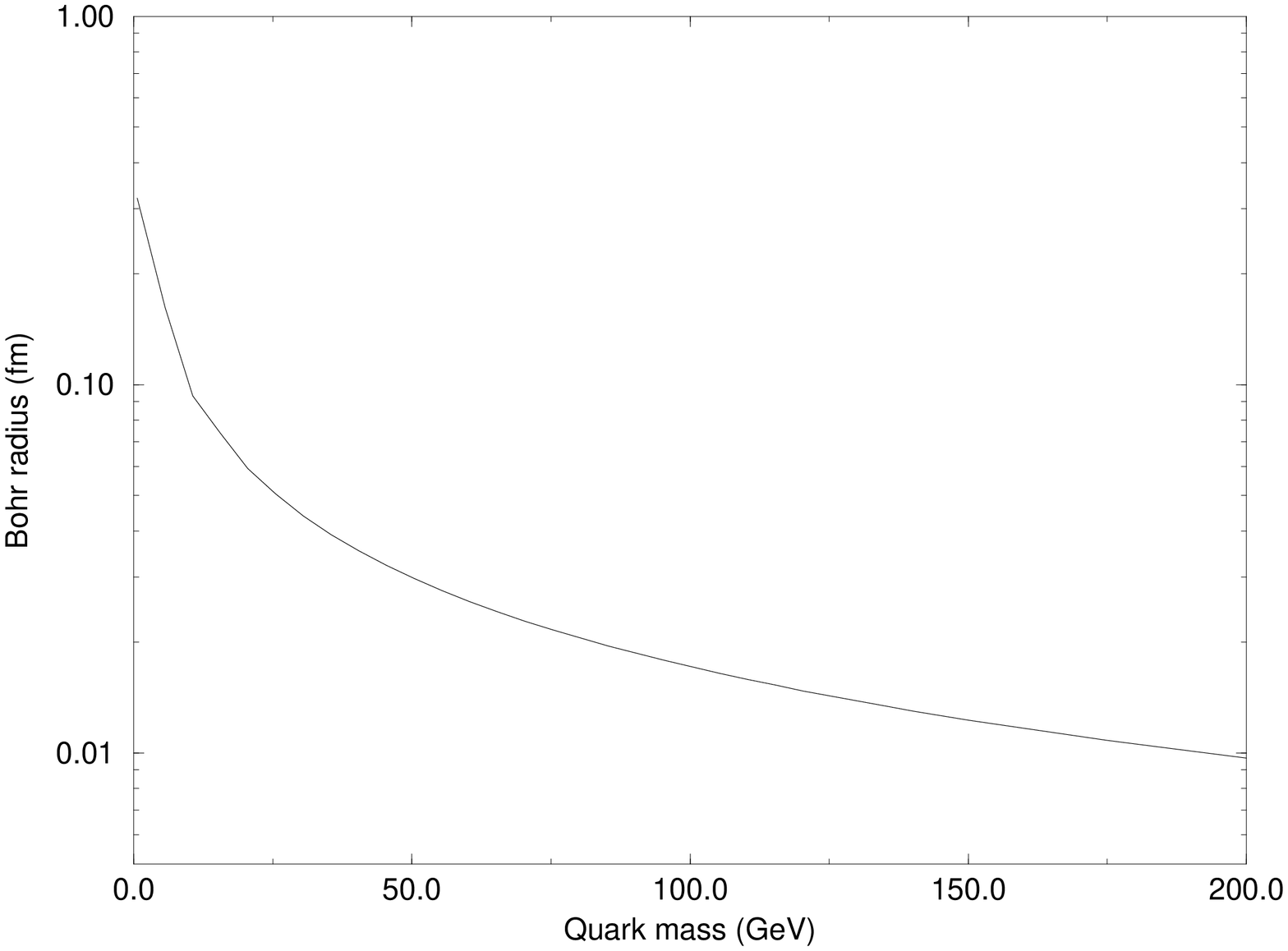,width=0.9\textwidth,bbllx=30pt,bblly=70pt,bburx=570pt,bbury=750pt,angle=270}}
\end{center}
\caption{{\it Bohr radius versus quark mass}}
\label{fig0}
\end{figure}

We shall use the two--loop expression for  $\alpha_s$ \cite{2LOOPS} ,

\begin{equation}
\alpha_s(Q^2)= \frac{4 \pi}{\beta_0 \log \left [ Q^2/\Lambda_{\overline{MS}} \right ] }
\left \{ 1-\frac{2 \beta_1}{\beta_0^2} \frac{\log \left [ \log \left [ Q^2/\Lambda_{\overline{MS}}
\right ] \right ]}
{\log \left [ Q^2/\Lambda_{\overline{MS}} \right ] }
 \right \}
\label{eq:2loop}
\end{equation}
with $ \beta_0=11-\frac{2}{3} \, n_f \;,\; \beta_1=51-\frac{19}{3} \, n_f$
evaluated at a fixed scale $Q^2=1/r^2_B$ \cite{NOI} , where $r_B$ is the Bohr radius 
\begin{equation}
r_B= \frac{3}{4 \mu \alpha_s}
\label{eq:rbohr}
\end{equation}
In figure (\ref{fig0}) we show the behavior of the Bohr radius as a function
of the quark mass for a quarkonia bound state under the effect of a Coulombic
interaction. The energy is given by

\begin{equation}
E_n = -\frac{8}{9} \; \frac{\mu \alpha_s^2}{n^2}
\label{eq:en}
\end{equation}
where $\mu $ is the reduced mass.
%

To estimate $t_R$ ,  we shall make use of the virial
theorem, which, for the Coulombic potential states that $ \langle T \rangle =
-\frac{1}{2} \langle V \rangle$
($T$ and $V$ are respectively the kinetic and potential energies). 
From the energy expression (\ref{eq:en}), 
we have $\langle v^2 \rangle = 8/9 \: \alpha_s^2$ , which leads to 
$t_R = \sqrt{2} \pi \: 9/(4 \alpha_s^2 m_t) $. 
As $\Gamma_{t \overline{t}} \approx 2cm_t^3$ (c is a constant) , this 
existence criterion allows for toponium formation if  
$\tau_{t \overline{t}} > t_R$, namely for 
$m_t < \sqrt{\sqrt{2}/(9\pi)} \; \alpha_s/\sqrt{c}$ .

Another slightly different criterion \cite{NOI} states that the formation of a 
hadron can occur only if the level splitting between the lowest lying levels of
the bound state, which depends upon the strength of the strong force between
the quarks and their relative distance \cite{FADIN}, is larger than the 
natural width of the state.
In this case, we should have a bound state if \mbox{ $ \Delta E_{2S-1S} \ge 
\Gamma_{t \overline{t}}$ . } From (\ref{eq:en}) we read the splitting of
the levels, $\Delta E_{2S-1S} = 4/3 m_t \alpha_s^2 $ .  In this case, 
the estimate for toponium existence is for values of top quark mass such that
$m_t< 1/\sqrt{6} \; \alpha_s/\sqrt{c}$ .

 A comparison with the previous method shows that the mass bound is larger 
for a factor $\sqrt{3\pi/(2\sqrt{2})}$ , that is approximately
two times larger, and we conclude that the energy level splitting criterion
is less stringent than the comparison of the revolution time. We decide 
therefore to employ the looser condition: if the formation is prohibited
by the $\Delta E_{2S-1S}$ criterion, then it is certainly prohibited 
also from the other method.

\begin{figure}[p]
\begin{center}
\leavevmode
\mbox{\epsfig{file=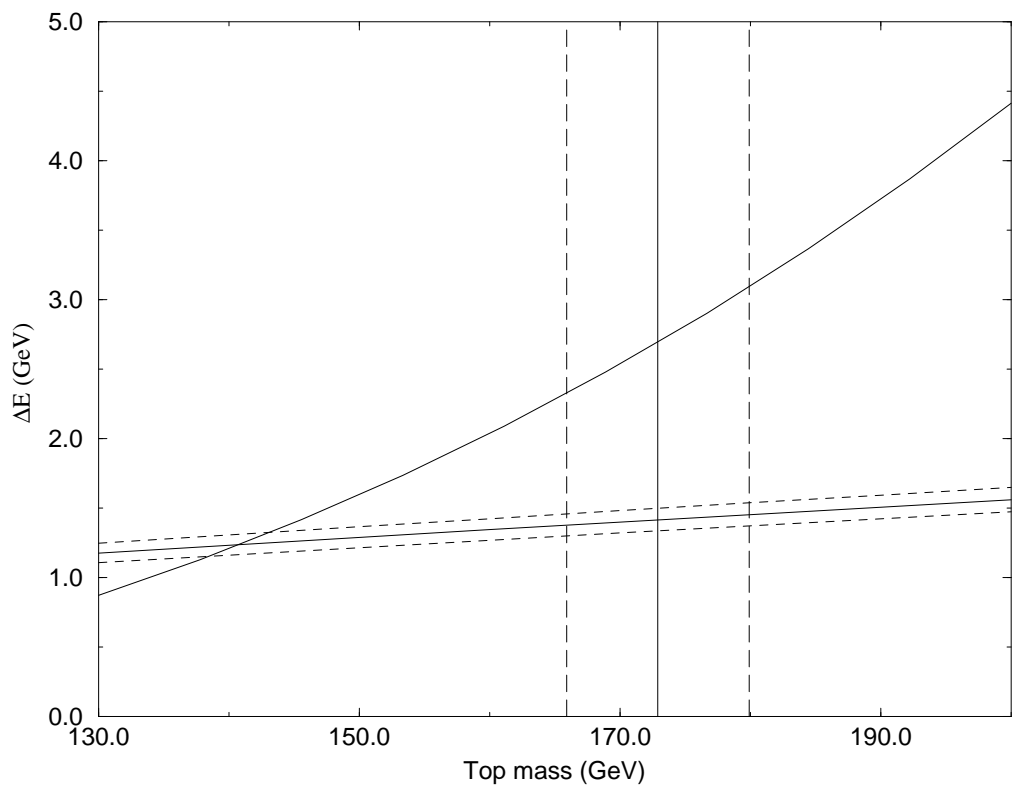,width=0.9\textwidth,bbllx=30pt,bblly=70pt,bburx=
570pt,bbury=750pt,angle=-90}} \end{center}
\caption{{\it Toponium width compared to energy separation of the Coulombic 
model. The stripe represents the indetermination of $\Delta E$ due to the 
value of $\alpha_s$ . The vertical bars
show the indetermination of the top mass.}}
\label{fig1}
\end{figure}

In  figure (\ref{fig1}), we show a comparison of the energy splitting and the
toponium width $\Gamma_{t \overline{t}} $ , as a function of the 
top mass. We shall use the value of  $\Lambda_{\overline{MS}}$ for 
which $\alpha_s(M_Z)=0.118$ \cite{PDG}, and we use for $\alpha_s$ the 
scale (\ref{eq:rbohr}).
From this figure, it is possible to see that for most recent top masses and the
average value for the top mass, $m_t=172.9 \pm 7.0 \; GeV$, toponium bound
state formation is rather unlikely. 

\section{T--Mesons from potential models}
For the T--mesons, the less stringent criterion of level splittings will be 
used. Notice that 
from \cite{NOI} the comparison of the Coulombic potential results with
some other models gave substantially similar results, thus confirming
the validity of this approach.

The problem for the T--meson, unlike toponium, is difficult due to the
 identification of the proper QCD scale for this problem in the 
Coulombic potential. The 
interquark force cannot clearly be set simply at the top mass scale, 
but it has to include both the information of the top and the light 
quark mass scale. In order to achieve
it, we use for the scale the inverse of the Bohr radius (\ref{eq:rbohr})
calculated in terms of the reduced mass of the T--meson, which in turn,  
because of the heavy top, is approximately the light quark mass.
In addition, we should also include calculations using 
other potentials, such as Martin's \cite{MARTIN}

\begin{equation}
V(R) = -8.064 + 6.8678 \; r^{0.1}
\label{eq:martin}
\end{equation}
(the units are in $GeV$), and the model of Grant, Rosner and Rynes
\cite{ROSNER}:

\begin{equation}
V(r) = \frac{\lambda}{\alpha} \; (r^{\alpha}-1) +c
\label{eq:rosner}
\end{equation}
These have the property of being independent of QCD coupling $\alpha_s$ , an 
important consideration since the energy scale considered is low, of 
the order of few fractions of GeV.
For quark masses we use the constituent one, whose approximate values 
are $m_b=5.0$, $m_c=1.5$, $m_s=0.5$, $m_u=m_d=0.3$ (in GeV). 
In table (1), we present a table of energy splitting values, for the 
bound states from the potential models. For sake of comparison we also include
the prediction of the Coulombic model and $\alpha_s$ values.
As for toponium, we shall use $\Lambda_{\overline{MS}}$ such that 
$\alpha_s(M_Z)=0.118 \; GeV $ \cite{PDG}, and the scale 
from (\ref{eq:rbohr})  .
\pagebreak
\par\vskip 10 mm
\centerline{\bf Energy splitting values $\Delta E$ for potential models (GeV)}
$$
\begin{array}{||c|c|c|c|c||} \hline
quark & Martin  & Rosner & Coulomb & \alpha_s \\ \hline
b & 0.56 & 0.65 & 0.29 & 0.30 \\ \hline
c & 0.60 & 0.61 & 0.20 & 0.45 \\ \hline
s & 0.31 & 0.41 & 0.20 & 0.78 \\ \hline
u,d & 0.08 & 0.02 & 0.24 & 1.09 \\ \hline
\end{array}
\label{table0}
$$
\vspace*{5mm}

{\bf Table 1:} {\sl  Values of $\Delta E$ from different potential models
(PM). We show  $\Delta E $ as a function of the light quark mass for 
the potentials given in formul\ae  (\ref{eq:martin}), (\ref{eq:rosner})
and (\ref{eq:coulomb}).  }

\par\vskip 10 mm
 We see that $\Delta E$ increases with the mass of the 
light quark, except for the Coulombic case, whose high $\alpha_s$ values
for light quarks make perturbative calculations unreliable.

In the following, we investigate the bound state system by means of a model
that includes relativistic corrections.
The t quark inside the $ t \overline{q} $ system moves nonrelativistically 
( $ v^2/c^2 \sim 0.01 $ for $ m_t \approx 173 $ GeV ). 
In order to show whether the lighter quarks need a relativistic treatment
we employ the Salpeter equation:

\begin{equation}
 \left [ \sqrt{- \nabla^2+m_t^2} + \sqrt{- \nabla^2+m_q^2} + V(r) 
\right ] \psi = E \psi
\label{eq:bs}
\end{equation}
with potentials from the nonrelativistic models.
Instead of dealing with the mathematical difficulties of finding the 
eigenvalues of (\ref{eq:bs}) because of the square root operator, we 
shall apply the Rayleigh--Ritz variational method with suitable trial
functions. This procedure has already been successfully applied for 
the $ B $ and $ D $ meson decay constants \cite{HWANG}. 
By means of a Fourier transform on (\ref{eq:bs}), we have

\begin{equation}
 \left [ \sqrt{p^2+m_t^2} + \sqrt{p^2+m_q^2} + V(r) 
\right ] \psi = E \psi
\label{eq:bs2}
\end{equation}

Writing explicitly the dependence of $ \psi $ upon an extremum 
parameter $ \xi $ , we find that the energy of the state is given by 
minimizing the expectation value of $ H $ in (\ref{eq:bs})

\begin{equation}
 \langle H \rangle \: = \: \langle \psi(\xi)|H|\psi(\xi) \rangle \: =  \: 
E(\xi) \; ;  \; \frac{dE(\xi)}{d \xi} = 0 \; \mbox{for} \; \xi = \overline{\xi}
\label{eq:minima}
\end{equation}
The energy levels depend on the number of nodes of the trial function:
no nodes for the ground state, and one node for the first excited level.
We have performed the calculation for two kinds of trial wavefunctions: 
hydrogen--type wavefunction, coming from a Coulombic potential, and Gaussian
wavefunction, from the harmonic oscillator, thus mimicking the short and
the long range behavior of the interquark force respectively.

For the $ 1S $ hydrogen--like function, we choose

\begin{equation}
  \psi(r) = \; \frac{1}{\sqrt{4 \pi}} \;  \frac{2}{a^{3/2}} \; e^{-r/a}
\label{eq:c1s}
\end{equation}
and its Fourier transform, which reads:

\begin{equation}
  \widehat{\psi}(p) = \; \frac{2 \sqrt{2}}{\pi} \; 
 \frac{a^{3/2}}{\left ( 1+a^2p^2 \right ) ^2} 
\label{eq:cf1s}
\end{equation}
$a$ being the variational parameter. The $ 2S $ function is:

\begin{equation}
  \psi(r) = \; \frac{1}{\sqrt{8 \pi}} \;  a^{-3/2} \;
\left (1- \frac{r}{2a} \right )  e^{-r/2a}
\label{eq:c2s}
\end{equation}
and

\begin{equation}
  \widehat{\psi}(p) = \; \frac{16}{\pi} \; a^{3/2} \; 
 \frac{(4a^2p^2-1)}{\left ( 1+a^2p^2 \right ) ^2} 
\label{eq:cf2s}
\end{equation}
respectively. For the Gaussian wavefunction, the $1S$ is:

\begin{equation}
  \psi(r) = \; \left (\frac{\mu}{\sqrt{ \pi}}\right )^{3/2} \; e^{-\mu^2 r^2/2}
\label{eq:g1s}
\end{equation}
and

\begin{equation}
  \widehat{\psi}(p) = \; \frac{1}{\left (\sqrt{ \pi} \mu \right )^{3/2}} \;
 e^{-p^2/2 \mu^2 }
\label{eq:gf1s}
\end{equation}
The $2S$ function is given by:

\begin{equation}
  \psi(r) = \; \sqrt{\frac{1}{(5 \sqrt{\pi}-8) \pi}} \; \mu^{3/2} \;
(1-r \mu) \; e^{-\mu^2 r^2/2}
\label{eq:g2s}
\end{equation}
and

\begin{equation}
 \widehat{\psi}(p) = \; \left [ \frac{2}{\sqrt{2 \pi}} -L^{1/2}_{1/2} 
\left ( \frac{p^2}{2 \mu^2} \right )
 \right ] \; \sqrt{\frac{1}{(5 \sqrt{\pi}-8) \pi}} \; \mu^{-3/2} 
\; e^{-p^2/2\mu^2}
\label{eq:gf2s}
\end{equation}
where $ \mu $ is a different minimization parameter from the previous one. 
$L_n^a(x)$ is the generalized Laguerre polynomial, obeying the differential
equation $ x y''+(a+1-x)y'+ny=0 $ and the orthogonality relation
$\int_0^{\infty} L_m^a(x) L_n^a(x) x^a e^{-x} dx = 0$ for $m \ne n$
(for $a=0$ , we retrieve the usual Laguerre polynomials).

Technically, one splits $H=T+V$ and then computes the average

\begin{equation}
\begin{array}{c}
  E(\xi)= \langle \widehat{\psi}(p)|T(p)|\widehat{\psi}(p \rangle +
 \langle \psi(r)|V(r)|\psi(r) \rangle = \\
 \\ 
4 \pi \int_0^{\infty} dp \: p^2 \: \widehat{\psi}(p) \left [ 
\sqrt{p^2+m_t^2}+\sqrt{p^2+m_q^2} \right ] \widehat{\psi}(p) + 
4 \pi \int_0^{\infty} dr \: r^2 \: \psi(r) V(r) \psi(r)
\end{array}
\label{eq:average1} 
\end{equation}
Below, we present a comparative table with results from both the 
potential models and two kinds of wavefunctions.
\par\vskip 10 mm
\centerline{\bf $ \Delta E$ for potential and relativistic models (GeV)}
$$
\begin{array} {||c|c|c|c|c|c|c||} \hline
\multicolumn{1}{||c|}{\mbox{}} & \multicolumn{2}{c|}{PM} &
\multicolumn{2}{c|}{RPM, Coulomb} & \multicolumn{2}{c||}{RPM, Gauss}
 \\ \hline
quark  & Martin & Rosner & Martin & Rosner & Martin & Rosner  \\ 
\hline
b & 0.56 & 0.65 & 0.47 & 0.57 & 0.53 & 0.61 \\ \hline
c & 0.60 & 0.61 & 0.45 & 0.39 & 0.55 & 0.35 \\ \hline
s & 0.31 & 0.41 & 0.41 & 0.09 & 0.44 & 0.00 \\ \hline
u,d & 0.08 & 0.02 & 0.38 & 0.01 & 0.34 & 0.00 \\ 
\hline
\end{array} 
\label{table01}
$$
\vspace*{5mm}

{\bf Table 2:} {\sl Comparison of $\Delta E$ results from potential models 
(PM) and relativistic potential models (RPM) for potentials 
(\ref{eq:martin}) and (\ref{eq:rosner}) . ``Coulomb'' and ``Gauss'' labels 
indicate the two different trial wavefunctions.}

\par\vskip 10 mm
As we should expect, major differences between the nonrelativistic
and relativistic potential models arise for the lighter (and hence faster) 
quarks. Also, the 
Martin potential is less sensitive to mass changes. We also notice that
there are no significant differences between different kinds of 
trial wavefunctions; 
confirming thus the reliability of this approach.

\section{Heavy Quark Theory approach}

 For performing these calculations, we shall now use a different method 
borrowed from heavy quark effective theory (HQET) . 
This has the advantage of giving results at a scale ($u$ and $d$ quarks)
for which some potential models predictions may not be reliable enough.
The Coulombic model is one of these, because $\alpha_s$ is of order 1 in this
 case.

In this model, a meson (hadron) 
containing a single heavy quark ($m_Q \gg \Lambda_{\overline{MS}}$) is
considered. The heavy quark's momentum can be written as

\begin{equation}
  p_Q=m_Q \cdot v + k
\label{eq:momentum}
\end{equation}
$k$ is the ``residual'' momentum, which measures the degree to which the quark 
is off--shell \cite{WISE}; $v$ is the velocity satisfying $v^2=1$ .
The quark $Q$ exchanges only small momenta with the rest of the hadron, so it
is essentially on shell, $p_Q^2=m_Q^2$ . $Q$ behaves like a static electric 
and chromomagnetic field source. The properties of the light degrees of freedom
do not depend upon the flavor and mass of the heavy field.
Since the top quark is very heavy, it plays the role of the heavy quark and 
we could apply HQET to give some information on the energy of the bound state.

The mass $M$ of the meson is here expressed as \cite{LUKE}

\begin{equation}
  M = m_Q + \overline{\Lambda} +  O \left ( \frac{1}{m_Q} \right )
\label{eq:hqmass}
\end{equation}
where $ \overline{\Lambda} $ is a positive contribution to $M$ . The 
``inertia'' parameter $ \overline{\Lambda} $ has no dependence on the heavy
degrees of freedom:
\begin{equation}
 \overline{\Lambda}=\overline{\Lambda}_q \equiv  \lim_{m_Q\rightarrow \infty}  
\left (M-m_Q \right )
\label{eq:lambdadef}
\end{equation}

On the other hand, from the potential models (PM) it is possible to write, for 
the meson mass,

\begin{equation}
  M = m_Q + m_q - E_b
\label{eq:pmmass}
\end{equation}
where $m_q$ is the mass of the light quark, $E_b \; (>0)$ is the binding energy
\cite{NOI} .
Comparing (\ref{eq:hqmass}) and (\ref{eq:pmmass}), we arrive at the 
identification

\begin{equation}
  \overline{\Lambda}_q = m_q - E_b
\label{eq:energy}
\end{equation}
neglecting terms of order $ O(1/m_Q^2)$ , an operation valid for
 \mbox{$m_t \approx 173 \; GeV$ .}
 We find therefore that the 
quantity $\overline{\Lambda}_q$ gives substantially the energy of the light
degrees of freedom\footnote{ A similar equivalence has 
been obtained in \cite{AHMADY} .}. In order to compute $\Delta E$, we need 
some estimates for $\overline{\Lambda}_q$ .

A rigorous lower bound on the value of $\overline{\Lambda}_q$ has been derived 
\cite{GURALNIK}, and leads to the values (in $GeV$)

\begin{equation}
\overline{\Lambda_q}(Q\overline{u}) \ge 0.057 \; , \; 
\overline{\Lambda_q}(Q\overline{d}) \ge 0.076 \; , \;
\overline{\Lambda_q}(Q\overline{s}) \ge 0.343
\label{eq:lval}
\end{equation}
for a meson formed by a heavy quark $Q$ and an $u, d$ and $s$ quark 
respectively. 
This lower bound has been obtained using the Euclidean path integral 
formulation of QCD. The Cauchy--Schwarz inequality has been used to derive
inequalities among Euclidean correlation functions, obtaining $m(\overline{Q}
\Gamma q)-m_q \ge 1/2 \: m'(\overline{q}i\gamma_5 q)$ . The first term of
the LHS is the heavy meson mass, while the RHS is the mass of the meson 
created by the light degrees of freedom, neglecting annihilation diagrams.

Equation (\ref{eq:hqmass}) gives us the binding energy of the 
lowest state, while we need the difference in energy. 
From the above
\begin{equation}
m_q -E_b = \overline{\Lambda}_q \ge \gamma_q
\label{eq:estim}
\end{equation}
where $\gamma_q$ is the lower bound on $\overline{\Lambda}_q$ . Moreover, we
have  $E_b \ge \Delta E_{2S-1S}$ , and we obtain from the lower
bound on $ \overline{\Lambda}_q $ , an upper bound for the quantity 
$\Delta E_{2S-1S}$ :

\begin{equation}
\Delta E_{2S-1S} \le m_q - \gamma_q
\label{eq:estimfin}
\end{equation}
From equation (\ref{eq:pmmass}) and constituent mass values for the light 
quarks, we obtain the following results:
\begin{equation}
\Delta E_{2S-1S}(t \overline{u}) \le 0.243 \; , \;
\Delta E_{2S-1S}(t \overline{d}) \le 0.224 \; , \;
\Delta E_{2S-1S}(t \overline{s}) \le 0.257
\label{eq:e1}
\end{equation}
with $m_u=m_d=0.3 \, GeV$ ; $m_s=0.5 \, GeV$ .

The values in equation (\ref{eq:e1}) have been derived using chiral 
perturbation theory,
and such a method is not applicable for heavier quarks. There is still the 
bound $ \overline{\Lambda}_q \ge \frac{1}{2} \, m'(\overline{q}i \gamma_5 q) $
\cite{GURALNIK} , and therefore $ \overline{\Lambda}_q \ge 0 $ .
With this less stringent bound and using the aforementioned values, we have 
(in $GeV$):

\begin{equation}
\Delta E_{2S-1S}(t \overline{c}) \le 1.5 \; , \;
\Delta E_{2S-1S}(t \overline{b}) \le 5.0
\label{eq:e2}
\end{equation}
 assuming $m_c= 1.5 \; GeV$, $m_b = 5.0 \; GeV $ .
We notice that
the estimate of the HQET for the $\Delta E_{2S-1S}$ for the heavier quarks is 
rather large due to a poor estimate on $\overline{\Lambda}_q$ , yet it is
consistent with the potential models. 

\section{Discussion and conclusions}
In table (3), we present a summary of the results which have to be compared
to the width of the T--meson. If $ \Delta E_{2S-1S} \ge 
\Gamma_{t\overline{q}}$ , then
there is an opportunity of formation for the bound state.
Since the quarks could decay in an independent manner from each other, 
one has $\Gamma_{t \overline{q}}=\Gamma_t + \Gamma_q$ . We know however 
that $\Gamma_q 
\approx m_q^5$ for a light quark, while $\Gamma_t \approx m_t^3$ , leading
to $\Gamma_t \gg \Gamma_q $, thus $\Gamma_{t \overline{q}}$ is essentially 
given by the top width.
\par\vskip 10 mm
\centerline{\bf Summary of $\Delta E$ results (GeV)}
$$
\begin{array} {||c||c|c|c||} \hline
quark & PM & RPM & HQET \\ \hline
b & 0.29-0.65 & 0.47-0.61  & 0.00-5.00 \\ \hline
c & 0.20-0.61 & 0.35-0.45  & 0.00-1.50 \\ \hline
s & 0.20-0.41 & 0.00-0.44  & 0.00-0.26\\ \hline
u,d & 0.02-0.24 & 0.00-0.38  & 0.00-0.24 \\ \hline
\end{array} 
\label{table1}
$$
\vspace*{5mm}

{\bf Table 3:} {\sl Summary results of $\Delta E$ ranges from potential
models (PM), relativistic models (RPM) and heavy quark effective 
theory estimates (HQET).}

\par\vskip 15mm

\begin{figure}[p]
\begin{center}
\leavevmode
\mbox{\epsfig{file=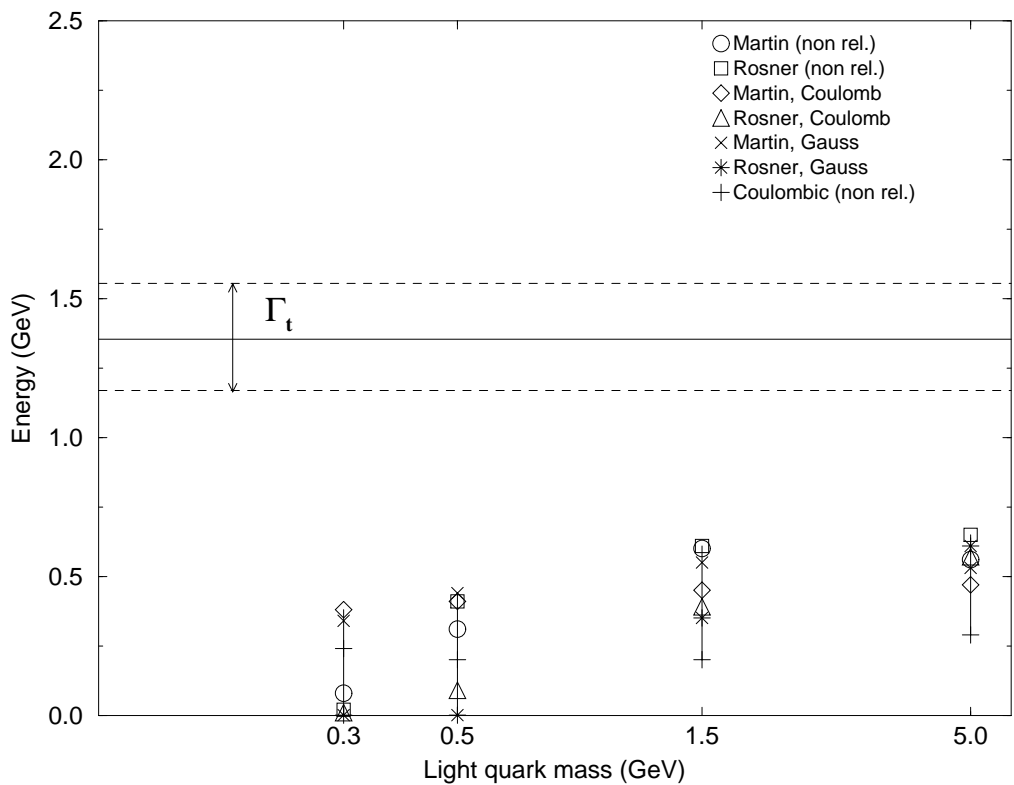,width=0.9\textwidth,bbllx=30pt,bblly=70pt,bburx=570pt,bbury=750pt,angle=270}}
\end{center}
\caption{{\it Energy splittings from non relativistic and 
relativistic potential models for different light quark masses, are compared
with the \mbox{T--meson} width (full and dashed horizontal lines).}}
\label{figure3}
\end{figure}

The tables (1), (2) and (3) have to be compared with the width values of the 
top. Considering the average value $m_t=172.9 \pm 7.0 \; GeV $ for the top 
mass, we have from (\ref{eq:width}) (in $GeV$):

\begin{equation}
\Gamma_t= 1.17 \; (m_t=165.9) ,\; \Gamma_t= 1.35 \; (m_t=172.9) ,\; \Gamma_t= 1.55 \; (m_t=179.9)
\label{eq:widval}
\end{equation}
Figure (\ref{figure3}) shows the window in energy ranges with respect to 
the decay width of the T--meson, which is drawn in full line. The dashed 
horizontal lines represent the indetermination due to the error on the 
top mass.

We see that the values for $\Gamma_t$ are larger than the one predicted 
from both the PM and the RPM. 
It must be noticed that some of the values from the HQET, the one for the 
heavier quarks (see table (3)),
seem to allow for \mbox{T--meson} formations, but it has to be stressed 
that the 
estimates for the $c$ and $b$ quarks are rather poor, since at the present
time a more precise estimate on $\overline{\Lambda}_{c,b}$ is lacking.

From a combination of these  calculational techniques we may conclude that
there is little evidence of possibility of formation for bound states of $t$ 
quark and lighter quarks.

\vspace{10mm}
{\bf Acknowledgments}
\vspace{5mm} 

The author wishes to thank G. Pancheri and Y. Srivastava for valuable help, 
stimulating discussion and careful reviewing of the manuscript.

\end{document}